\documentclass[prb,amsmath,amssymb]{revtex4}

\usepackage{color}

\newcommand{\bn}[1]{\mbox{\boldmath $#1$}}
\def\e{\mathop{\rm \mbox{{\Large e}}}\nolimits}

\def\ni{\noindent}
\newcommand{\be}{\begin{equation}}
\newcommand{\ee}{\end{equation}}
\newcommand{\ba}{\begin{array}}
\newcommand{\ea}{\end{array}}
\newcommand{\bea}{\begin{eqnarray}}
\newcommand{\eea}{\end{eqnarray}}
\newcommand{\beann}{\begin{eqnarray*}}
\newcommand{\eeann}{\end{eqnarray*}}
\newcommand{\mb}{\mbox}
\newcommand{\bm}{\boldmath}

\newcommand{\bna}[1]{\mbox{\boldmath $#1$}}

\begin{document}

\title[Symmetries Requirements for Transfer Matrices]{Symmetries and General Principies in the
Multiband Effective Mass Theory: A Transfer Matrix Study.}

\author{L. Diago-Cisneros$^{\dag}$\footnote{permanent address: Dpto. de F\'{\i}sica Aplicada,
Facultad de F\'{\i}sica, Universidad de La Habana, C.P.10400, La Habana,Cuba.}, H.
Rodr\'{\i}guez-Coppola$^{\ddag}$, R. P\'{e}rez-\'{A}lvarez$^{\S}$ and P. Pereyra$^{\dag}$}

\affiliation{\dag\ Departamento de Ciencias B\'{a}sicas, UAM-Azcapotzalco, C.P.02200 D.F.
M\'{e}xico.}

\affiliation{\ddag\ Dpto. de F\'{\i}sica Aplicada, Facultad de F\'{\i}s., Univ. de La
Habana, C.P.10400, La Habana,Cuba.}

\affiliation{\S\ Dpto. de F\'{\i}sica Te\'orica, Facultad de F\'{\i}sica, Univ. de La
Habana, C.P.10400, La Habana,Cuba.}

\date{\today}

\begin{abstract}

We study the time reversal and space inversion symmetry properties of those transfer matrices
mostly used in the calculation of energy spectra and transport-process quantities. We determine the
unitary transformation relating transfer matrices. We consider the Kohn-L\"uttinger model for a
\textit{quasi}-$2$D system and show that even though the system studied in the $(4\times 4)$ scheme
satisfies all the symmetry requirements, the $(2\times 2)$ subspaces do not fulfill such
constrains, except in the $\Gamma$ point of the Brillouin Zone. We find new exchange properties
between the $(2\times 2)$ subspace quantities.
\end{abstract}

\maketitle

%\submitto{\Phys Scripta}

\section{Introduction}

In the last years a large number of theoretical studies of multilayered semiconductor structures,
using different schemes of $\mbox{\boldmath $k$}\cdot \mbox{\boldmath $p$}$ multiband Hamiltonians,
have been published \cite
{APB,OW,SSG,SN,Broido85,SC1,CS1,EAP,IM,IMT,SP95,GF1,LCR00,GF2,GF3,Mireles99,Radhia03}. Besides the
extensive application of these models, the symmetry problem in general, and the specific
characteristics in particular, such as the time reversal invariance in the Kramer $(2\times 2)$
subspaces of the Kohn-L\"{u}ttinger (KL) model and the parity of heavy- (\textit{hh}) and
light-hole (\textit{lh}) states, have been only partially considered in a small number of
references\cite{APB,GF1,LCR00,GF2,GF3}. The increasing interest in studying transport properties
within the $\mbox{\boldmath $k$}\cdot \mbox{\boldmath $p$}$ approximation puts forward the need of
a simple procedure to obtain scattering amplitudes from Envelope Function Approximation (EFA)
models, and to establish clear requirements to preserve the fundamental physical symmetries. Many
physical properties, such as tunnelling and quantum coherence, depend strongly on the symmetries
preserved by the system-particle interaction\cite{Shanhui99}. Recently, remarkable efforts have
been done to study the breaking of discrete symmetries \cite
{GF1,LCR00,PFF97,ME97,S-P97,Y-S97,TT97}. However some issues remain to be studied. The lack of a
theoretical analysis of whether the time reversal symmetry  and the space inversion symmetry in the
Kramer $(2\times 2)$ subspaces are fulfilled or not, shows the need of a detailed and comprehensive
analysis of symmetry problems in the KL and other EFA models. This is the aim of the present paper.

An important quantity, where the original Hamiltonian symmetries can be checked out in a simple way
is the transfer matrix (TM). This object is being used with increasing frequency to study transport
properties of different type of systems and within quite distinct approaches. Basically, two types
of transfer matrices are being used in the literature. One of these transfer matrices, called here
of the \textit{first kind}, relates a field $\mbox{\boldmath $F$}(z)$ and its derivative
$\mbox{\boldmath $F$}^{\prime }(z)$ at any two points $z_{1}$ and $z_{2}$ of the scatterer
system\cite {James49,Erdos82,GMolinerVelasco92}. Another transfer matrix, called here of the
\textit{second kind}, connects state vectors $\bn{\Phi}(z)$ that are expressed in the propagating
modes representation \cite {Luttinger51,Borland61,MPK88,GMolinerVelasco92,PPP95,PPP98a,PPP98b}.
Because of its simple functional relation with the scattering amplitudes, the latter matrices were
mainly used in the scattering approaches to deal with transport processes. The rigorous fulfillment
of symmetry requirements, such as time reversal invariance (TRI), spin rotation (SR), space
inversion invariance (SII) and flux conservation (FC) principle, were amply discussed in the
literature in connection with these transfer matrices, both for $1$-D one-channel and for $3$-D
multichannel systems \cite {Borland61,MPK88,PPP95,PPP98a}. Transfer matrices of the first kind or
their \textit{associated transfer matrices} (which relate $\bn{F}(z)$ and a certain linear form
$\bn{G}(z)= \bn{v}(z)\bn{F}^{\prime}(z) + \bn{u}(z)\bn{F}(z)$, at any two points $z_{1},$ and
$z_{2}$), were mainly used in connection with the solution of EFA models\cite{HRC96,RPA01c}. The
matrices $\bn{u}(z)$ and $\bn{v}(z)$ appear as coefficients in the matrix equation of motion
\cite{Malik99,RPA01c}. To obtain the transfer matrix in the EFA one needs to build up the vector
$\mbox{\bm $F$}(z)$, called the \emph{envelope }(see Refs.
\onlinecite{LCR00,GMolinerVelasco92,HRC96,Monsivais95,Griffiths01,Sheng96a,Sheng96b,RPA-JRS-00} and
references therein). TM of the first kind have been widely used to study the hole spectrum in
III-V, II-VI and IV semiconductors\cite
{Bastard89,Luttinger55,Kohn55a,Kohn55b,Kohn55c,Geller93a,Geller93b}.

Since the multiband KL model in the $\mbox{\boldmath $k$}\cdot\mbox{\boldmath $p$}$ approximation
is amply used in the current solid state physics, and the understanding of the transport properties
in heterostructure systems become of great interest, it is convenient to establish a clear bridge
between the first and the second kind of transfer matrices. To this purpose the analysis of the
symmetry requirements on the transfer matrices of the first kind and their relation with the
scattering amplitudes and transfer matrices of the second kind, is very much called for. We apply
our results to study with clear advantages and transparency the multichannel transport of heavy and
light holes through III-V semiconductor heterostructures \cite{LPRC02,LPCR04}. However the results
that we present here are to some extent also applicable to any EFA model.

Assuming the existence of $N$-coupled-differential-equations solutions, we define in Section
\ref{Def} the transfer matrices of the first and second kind and determine the unitary
transformation between them. In Section \ref{Sym}, we derive the mathematical conditions that the
transfer matrices of the first and second kind have to satisfy to fulfill the physical symmetries,
and we also recall those for the TM of the second kind that are known. We focus our attention on
the time reversal invariance, space inversion symmetry, and the flux conservation principle. In
Section \ref{KL} we consider the $(4\times 4)$ KL Hamiltonian and derive an explicit representation
of the unitary transformation determined in Section \ref{Def}. Next, we discuss the consequences of
the various symmetry requirements obtained in Section \ref{Def} on the Hamiltonians and related
quantities of the KL $(2\times 2)$ subspaces. Finally we present some concluding remarks and
comments.

\section{Transfer matrices: An outline of basic definitions}
 \label{Def}

In this Section, we will outline the well known transfer matrix definitions. Based on
these definitions, we will establish the relation between the transfer matrices of the
first and second kind.

Solving the system of equations associated to an EFA Hamiltonian, one obtains a set of
linearly independent eigenvectors. It is often convenient to consider, as will be the case
here, an orthogonal set of eigenvectors. With these eigenvectors we can build the envelope
function
\begin{equation}
\mbox{\boldmath $F$}(z)={\textstyle\sum\limits_{j=1}^{2N}}\,a_{i}\,%
\mbox{\boldmath $f$}_{i}(z),
\end{equation}
where $N$ is the dimension of the system of coupled differential equations, and the vector
\begin{equation} \label{FF}
\mbox{\boldmath $\Psi$}(z)=\left[
\begin{array}{c}
\mbox{\boldmath $F$}(z) \\
\mbox{\boldmath $F$}^{\prime}(z)
\end{array}
\right] .
\end{equation}

Alternatively, we can write the eigenvectors in the propagating
modes representation as
\begin{eqnarray}
 \label{LI}
\mbox{\bm $f$}_{j}(z) = \left(
\begin{array}{c}
g_{1j} \\
g_{2j} \\
... \\
g_{Nj}
\end{array}
\right) \; \mathop{\rm \mbox{{\Large
e}}}\nolimits^{iq_{j}z}=\mbox{\bm $g$}_j\mathop{\rm \mbox{{\Large
e}}}\nolimits^{iq_{j}z}
\end{eqnarray}
and define, in the plane-wave basis, the state vectors
\begin{equation}
 \label{SV1}
 \mbox{\boldmath $\Phi$}(z)=\left[
 \begin{array}{c}
 \mbox{\boldmath $a$}\stackrel{\rightarrow }{\mbox{\boldmath
 $\varphi$}}(z)
\\
 \mbox{\boldmath $b$}\stackrel{\leftarrow }{\mbox{\boldmath
 $\varphi$}}(z)
 \end{array}
 \right]
\end{equation}
where $\stackrel{\rightarrow }{\mbox{\boldmath $\varphi$}}(z)$ and $\stackrel{\leftarrow
}{\mbox{\boldmath $\varphi$}}(z)$ represent the right and left moving modes (when $q_{j}$ is real).
Both of them are $N$-dimensional vectors built in terms of the right- and left-travelling wave
functions $\stackrel{\rightarrow }{\varphi }_{i}\left( z\right) $ and $\stackrel{ \leftarrow
}{\varphi }_{i}\left( z\right)$ in the $i$-\textit{th} channel or propagating mode.

It is easy to see that, writing $\mbox{\boldmath $\Phi$}(z)$ as
\begin{equation}
 \label{SV2}
 \mbox{\boldmath $\Phi$}(z)=\left(
 \begin{array}{cc}
 \stackrel{\rightarrow }{\mbox{\boldmath $\varphi$}} & 0 \\
 0 & \stackrel{\leftarrow }{\mbox{\boldmath $\varphi$}}
 \end{array}
 \right) \left(
 \begin{array}{c}
 \mbox{\boldmath $a$} \\
 \mbox{\boldmath $b$}
 \end{array}
 \right)
\end{equation}
\noindent where $\mbox{\boldmath $a$}$ and $\mbox{\boldmath $b$}$
are $N$ dimensional vectors, and
\begin{equation}
\stackrel{\rightarrow }{\varphi}(z)= \left(
\begin{array}{cccc}
\stackrel{\rightarrow }{\mbox{\boldmath $\varphi$}}_{1}\left(
z\right) & 0 &
... & 0 \\
0 & \stackrel{\rightarrow }{\varphi }_{2}\left( z\right) & ... & 0 \\
... & ... & ... & ... \\
0 & 0 & .. & \stackrel{\rightarrow }{\varphi }_{\mb{\tiny N}}\left( z\right)
\end{array}
\right),
\end{equation}
\noindent a simple and obvious unitary transformation can be established between
$\mbox{\boldmath$\Psi$}(z)$ and $\mbox{\boldmath $\Phi$}(z)$. Since these vectors are used to
define the transfer matrices of the first and second kind, one can determine such unitary
transformation, i.e. \be
 \label{Psi-Phi}
 \bn{\Psi}(z)=\bn{\cal N}\bn{\Phi}(z)
\ee \noindent which is fundamental for the purpose of this paper. The specific structure of
$\mbox{\boldmath$\mathcal{N}$}$ depends on the particular Hamiltonian.

\subsection{Transfer matrices of the first kind}
Using the function $\mbox{\boldmath $\Psi$}(z)$ it is common to define the transfer matrix of the first kind
$\mbox{\boldmath $M$}_{fd}$ such that
\begin{equation}
 \label{DefMfd}
  \mbox{\boldmath $\Psi$}(z_{2})=\mbox{\boldmath $M$}_{fd}(z_{2},z_{1})%
  \mbox{\boldmath $\Psi$}(z_{1}).
\end{equation}

A slightly different definition for this type of transfer matrix is the associated
transfer matrix $\mbox{\boldmath $T$}$ \cite{RPA01c} defined by
\begin{equation}
 \label{DefT}
 \bn{\psi}(z_{2}) = \mbox{\boldmath $T$}(z_{2},z_{1})
 \bn{\psi}(z_{1})
\end{equation}
\noindent where
\begin{equation}
 \label{F&FL}
 \bn{\psi}(z)=\left[
  \begin{array}{c}
   \mbox{\boldmath $F$}(z) \\
   \mbox{\boldmath $G$}(z)
  \end{array}
  \right] \equiv \left[
  \begin{array}{c}
   \mbox{\boldmath $F$}(z) \\
   \mbox{\boldmath $v$}(z)%
   \mbox{\boldmath $F$}^{\prime }(z)+\mbox{\boldmath $u$}(z)\mbox{\boldmath $F$}(z)
  \end{array}\right].
\end{equation}
These two matrices are the first kind of matrices we use.
\subsection{Transfer matrices of the second kind}
Another TM that has been widely used in the literature that keeps a simple relation with
the scattering amplitudes is the matrix defined by \cite{MPK88}
\begin{equation}
\label{DefMsv}
\mbox{\boldmath $\Phi$}(z_{2})=\mbox{\boldmath $M$}_{sv}(z_{2},z_{1})%
\mbox{\boldmath $\Phi$}(z_{1}).
\end{equation}

As for the first kind of transfer matrices, there is also a slightly different definition
of the transfer matrix, usually called the coefficients transfer matrix defined by
\begin{equation}
\left(
\begin{array}{c}
\mbox{\boldmath $c$} \\
\mbox{\boldmath $d$}
\end{array}
\right) = \mbox{\boldmath $Q$}(z_{2},z_{1}) \left(
\begin{array}{c}
\mbox{\boldmath $a$} \\
\mbox{\boldmath $b$}
\end{array}
\right).
\end{equation}
Based on the previous definitions and the transformation (\ref{Psi-Phi}) it is clear that
both types of transfer matrices are related by the system-dependent unitary transformation
\begin{equation}
 \label{Msv-Mfd}
  \mbox{\boldmath $M$}_{sv}(z_{\mb{\tiny R}},z_{\mb{\tiny L}})=
  \mbox{\boldmath $\mathcal{N}$}^{-1}
  \mbox{\boldmath $M$}_{fd}(z_{\mb{\tiny R}},z_{\mb{\tiny L}})
  \mbox{\boldmath $\mathcal{N}$}.
\end{equation}
This is an important result. Based in the previous unitary transformation it is easy to obtain simple
functional relations with the scattering amplitudes. The transformation matrix
$\bn{\mathcal{N}}$, say
\be
 \label{Ncal}
  \bn{\mathcal{N}}= \left[ \ba{cccc}
  \bn{g}_{1} & \bn{g}_{2} & \;\;...\;\; & \bn{g}_{2N} \\
  d_{1}\bn{g}_{1} & d_{2}\bn{g}_{2} & \;\;...\;\; & d_{2N}\bn{g}_{2N}
  \ea \right],
\ee can be obtained when each linearly independent (LI) solution is written as a $(N\times 1)$
spinor, with no coordinate dependence (represented here by $\bn{g}_{j}$), times a plane wave. By
$d_{j}$ we denote the coefficient of $z$ in the exponent of the plane waves. In Section \ref{KL} we
will consider a particular KL Hamiltonian and we will obtain specific expressions for
$\mbox{\boldmath $M$}_{fd}$ and $\mbox{\boldmath $M$}_{sv}$ and the unitary transformation in
(\ref{Msv-Mfd}).

\section{ General FC, TRI and SII Requirements on M$_{fd}$ and M$_{sv}$}
 \label{Sym}

Let us now recall and deduce the constrains imposed on the transfer matrices by the
fulfillment of flux conservation and the physical symmetries of time reversal and space
inversion. For the sake of brevity, we report the most important relations and quote just
some of the necessary and well known results.

\subsection{Flux Conservation}

To obtain the flux conservation requirements we need to write the current density
\begin{equation}
j(z)=-\frac{i\hbar}{\,\,2m^{*}}\left[\mbox{$\Phi$}^{\dag}\mbox{\boldmath
$\nabla$}\mbox{$\Phi$}-\mbox{$\Phi$}^{T}\mbox{\boldmath $\nabla$}\mbox{$\Phi$}^{*}
\right],
\end{equation}
in terms of any of the previously introduced $2N$-dimensional spinors. The method of
derivation follows the usual calculation of the particle current in quantum mechanics. We
thus have
\be
 \label{fluxNband}
  j(z)=i\left[\bn{F}^{\dag}\bn{G}-\bn{G}^{\dag}\bn{F}\right].
\ee

It is found useful to emphasize that in the KL model this formula reduces to
$j(z)=2\mb{Im}[\bn{F}\hspace{0.3mm}'^{\dag}\bn{v}\bn{F}] -
2\bn{F}^{\dag}\bn{u}^{\dag}\bn{F}$. In other interesting cases as the Schr\"{o}dinger
equation and the $1$-D one-channel and for $3$-D multichannel systems \cite
{Borland61,MPK88,PPP95,PPP98a} it reduces to the known expression
$j(z)=\bn{F}^{\dag}\bn{F}' - \bn{F}'^{\dag}\bn{F}$.

When flux is conserved, which means $j(z_2) = j(z_1) $ at any two points, we immediately
obtain the identity:
\begin{equation}
 \label{FC4T}
  \bn{R}^{\dag}(z)\bn{\Sigma}_{y}\bn{R}(z)= \bn{M}_{fd}^{\dag }(z,z_{0})
  \bn{R}^{\dag}(z_{0})\bn{\Sigma}_{y}\bn{R}(z_{0})\bn{M}_{fd}(z,z_{0}),
\end{equation}
\ni where
$$
 \bn{R}(z) =
 \left\vert \ba{cc}
 \bn{I}_{\mb{\tiny N}} & \bn{O}_{\mb{\tiny N}} \\
 \,\, &  \,\,    \\
 \bn{u}(z) & \bn{v}(z)
 \ea \right\vert,
 $$
\ni and
$$
\mbox{\boldmath $\Sigma$} _{y}=\left[
\begin{array}{cc}
 \mbox{\boldmath $O$}_{\mb{\tiny N}} & -i\mbox{\boldmath $I$}_{\mb{\tiny N}} \\
 i\mbox{\boldmath $I$}_{\mb{\tiny N}} & \mbox{\boldmath $O$}_{\mb{\tiny N}}
\end{array}\right].
$$

Henceforth $\mbox{\boldmath $O$}_{\mb{\tiny N}}$/$\mbox{\boldmath $I$}_{\mb{\tiny N}}$ is the
corresponding $(N\times N)$ null/identity matrix. In a completely similar way, one obtains the FC
requirement for other transfer matrices. For the transfer matrices of the second kind the current
conservation implies
\begin{equation}
\mbox{\boldmath $d$}\mbox{\boldmath $c$}^{*}-\mbox{\boldmath $c$}
\mbox{\boldmath $d$}^{*}= \mbox{\boldmath $b$}\mbox{\boldmath
$a$}^{*}-\mbox{\boldmath $a$}\mbox{\boldmath $b$}^{*}
\end{equation}
\noindent which, using the transfer matrix definition, leads to the well known relation
\begin{equation}  \label{FC4Msv}
\mbox{\boldmath $\Sigma$}_{z}= \mbox{\boldmath $M$}_{sv}^{\dag }(z,z_{0})%
\mbox{\boldmath $\Sigma$}_{z}\mbox{\boldmath $M$}_{sv}(z,z_{0}),
\end{equation}
\noindent where $\bn{\Sigma}_{z}=\bn{\sigma}_{z}\bigotimes \bn{I}_N$, is the enlarged $(2N
\times 2N)$ Pauli $\sigma_z$ matrix (see Appendix \ref{MatTran}). This condition has been
explicitly deduced in various references\cite {Borland61,MPK88,PPP95}.

\subsection{Time Reversal Invariance}

TRI implies that the envelope function fulfills the relation
\begin{equation}  \label{OpTRI}
\mbox{\boldmath $K$}\hat{C}\mbox{\boldmath $F$}(z)=\mbox{\boldmath $F$}(z)\;,
\end{equation}
where the usual notation for the complex-conjugation operator $\hat{C}$ is
used, and $\mbox{\boldmath $K$}$ is a model-dependent matrix \cite
{GMolinerVelasco92,Cohen}. Using the definition (\ref{DefMfd}) in equation (%
\ref{OpTRI}) we obtain
\begin{equation}  \label{TRI4Mfd}
\mbox{\boldmath $M$}_{fd}(z,z_{0}) = \mbox{\boldmath $\Sigma$} %
\mbox{\boldmath $M$}_{fd}^{\ast }(z,z_{0})\mbox{\boldmath $\Sigma$}^{-1}
\end{equation}
with
\[
\mbox{\boldmath $\Sigma$} =\left[
\begin{array}{cc}
\mbox{\boldmath $K$} & \mbox{\boldmath $O$}_4 \\
\mbox{\boldmath $O$}_4 & \mbox{\boldmath $K$}
\end{array}
\right].
\]
$\mbox{\boldmath $T$}$, $\mbox{\boldmath $M$}%
_{sv}$ and $\mbox{\boldmath $Q$}$ fulfill similar identities. In the case of the matrix
$\mbox{\boldmath $T$}$, we obtain
\begin{equation}
\mbox{\boldmath $T$}(z,z_{0}) = \mbox{\boldmath $\Sigma$}
\mbox{\boldmath
$T$}^{\ast }(z,z_{0}) \mbox{\boldmath $\Sigma$}^{-1}.
\end{equation}
\noindent Concerning the transfer matrix of the second kind, it is well
known that the TRI condition for systems with spin-independent interactions
takes the form
\begin{equation}  \label{TRI4Msv}
\mbox{\boldmath $M$}_{sv}(z,z_{0}) = \mbox{\boldmath $\Sigma$}_{x} %
\mbox{\boldmath $M$}_{sv}^{\ast }(z,z_{0})\mbox{\boldmath $\Sigma$}_{x}^{-1},
\end{equation}
\noindent while for systems with spin-dependent interactions, this relation changes
slightly. The matrix $\mbox{\boldmath $\Sigma$}_{x}$ is given in the Appendix
\ref{MatTran}. In the particular case of systems of spin $1/2$ (see
Ref.\onlinecite{PPP95}), TRI implies the condition
\begin{equation}
\mbox{\boldmath $M$}_{sv}(z,z_{0})= \mbox{\boldmath $K$}\mbox{\boldmath $M$}%
_{sv}^{\ast }(z,z_{0})\mbox{\boldmath $K$}^{-1}.
\end{equation}

\subsection{Space Inversion}

Let us now assume that the system possesses a space inversion symmetry with respect to the
origin. It is convenient to emphasize that the space inversion we are interested here is
the transformation that changes only the sign to the coordinate perpendicular to the
interfaces of the Q$2$D system, leaving the signs of the in-plane wave vector components
$k_{x}$ and $k_{y}$ unchanged. If $\hat{S}_{\mb{\tiny I}}$ denotes such space inversion
operator, the space inversion invariance implies
\begin{equation}
 \label{OpSI}
 \hat{S}_{\mb{\tiny I}}\mbox{\boldmath $F$}(z)=p\mbox{\boldmath $s$}%
 \mbox{\boldmath $F$}(-z)
\end{equation}
where $p=\pm 1$ and $\mbox{\boldmath $s$}$ is a model-dependent matrix.

As the derivatives change the parity of the functions, the action of the space inversion operator
on the bi-field $ \mbox{\boldmath $\Psi$}(z)$ becomes
\begin{equation}  \label{OpSI-2field}
\hat{S}_{\mb{\tiny I}}\mbox{\boldmath $\psi$}(z)=p\mbox{\boldmath $S$}%
\mbox{\boldmath
$\psi$}(-z),
\end{equation}
with
\begin{equation}
\mbox{\boldmath $S$}= \left[
\begin{array}{cc}
\mbox{\boldmath $s$} & \mbox{\boldmath $O$}_4 \\
\mbox{\boldmath $O$}_4 & -\mbox{\boldmath $s$}
\end{array}
\right].
\end{equation}
\noindent Therefore, the SII requirement on the transfer matrix $\mbox{\boldmath
$M$}_{fd}$ is
\begin{equation}  \label{SI4Mfd}
\mbox{\boldmath $M$}_{fd}(z,z_{0})= \mbox{\boldmath $S$}^{-1}%
\mbox{\boldmath
$M$}_{fd}(-z,-z_{0})\mbox{\boldmath $S$}.
\end{equation}
\noindent Similar identities can be obtained for all the transfer matrices.
In particular, for $\mbox{\boldmath $M$}_{sv}$ we obtain:
\begin{equation}
\mbox{\boldmath $M$}_{sv}(z,z_{0})= \mbox{\boldmath $S$}_{sv}^{-1}%
\mbox{\boldmath $M$}_{sv}(-z,-z_{0})\mbox{\boldmath $S$}_{sv}.
\end{equation}
\noindent In the Appendix \ref{MatTran}, the explicit expression for $%
\mbox{\boldmath $S$}_{sv}$ is given.

\section{TM properties and Symmetry Requirements in the Kohn-L\"{u}ttinger Hamiltonian}
 \label{KL}

The first part of this section is devoted to obtain an explicit form for $\bn{\cal{N}}$ in the case
of the $(4\times 4)$ KL model. In the second part we will focus our attention on the symmetries of
the $(2\times 2)$ subspaces of this model. For the sake of completeness we shall briefly recall
some important and well-known results within the $(4 \times 4)$ KL
model\cite{APB,GF1,LCR00,GF2,GF3,Bass75,Bir72} and we will describe explicitly the consequences of
its TRI and SII symmetries on the $(2\times 2)$ subspaces. These consequences have not received
sufficient attention, however, because of their current interest concerning the elastic and
assisted transmission studied within the $(2\times 2)$ subspaces, deserve further clarification. We
shall discuss these implications for the cases in which the wave vector $\kappa$ (defined in the
interface planes) is either equal or different from $0$.

\subsection{Derivation of the transformation matrix \bn{\cal N}
            in the $(4 \times 4)$ KL model}

Dealing with the $(4 \times 4)$ KL model Hamiltonian it is usual to block-diagonalize it
\cite{Broido85} and to work then with $(2\times 2)$ instead of $(4\times 4)$ Hamiltonians. In this
way, the mathematical difficulties are highly simplified. This method remains, nevertheless, a very
useful tool to understand many of the intriguing physical properties of \textit{hh} and \textit{lh}
valence bands near the band edge, and we use it for the derivation of the transformations
(\ref{Psi-Phi}) and (\ref{Msv-Mfd}) in the $(4 \times 4)$ space as our main objective.

In the block-diagonalizing procedure, a unitary transformation $\mbox{\boldmath $U$}$ performs the
splitting of the original $(4 \times 4)$ Hamiltonian $\hat{\bn{H}}$ into two $(2 \times 2) $
blocks, which are labelled ``up''(\textit{u}) and ``low''(\textit{l}) \be
 \label{block}
 \bn{U}\hat{\bn{H}}\bn{U}^\dag\bn{U}\bn{F}(z) = \left[
 \ba{cc}
  \hat{\bn{\cal H}}_{u} & \bn{O}_{2} \\
  \bn{O}_{2} & \hat{\bn{\cal H}}_{l}
 \ea \right] \left[
  \ba{c} \bn{\cal F}_{u}(z) \\
   \bn{\cal F}_{l}(z)
  \ea \right] = E\left[
  \ba{c} \bn{\cal F}_{u}(z) \\
   \bn{\cal F}_{l}(z)
  \ea \right],
\ee
\ni where the blocks are given by
\be
 \label{Hu}
  \hat{\bn{\cal H}}_{u} = \left[
  \begin{array}{cc}
   A_{1}\kappa^{2}+B_{2}\hat{k}_{z}^{2}+V(z) & C_{xy}-iD_{xy}\hat{k}_{z} \\
   C_{xy}+iD_{xy}\hat{k}_{z} & A_{2}\kappa^{2}+B_{1}\hat{k}_{z}^{2}+V(z)
  \end{array}\right],\\
\ee
\be
 \label{Hl}
  \hat{\bn{\cal H}}_{l} = \left[
  \begin{array}{cc}
   A_{2}\kappa^{2}+B_{1}\hat{k}_{z}^{2}+V(z) & C_{xy}-iD_{xy}\hat{k}_{z} \\
   C_{xy}+iD_{xy}\hat{k}_{z} & A_{1}\kappa^{2}+B_{2}\hat{k}_{z}^{2}+V(z)
  \end{array}\right].
\ee
\noindent For the definitions of $A_{1}$, $A_{2}$, $B_{1}$, $B_{2}$, $C_{xy}$%
, $D_{xy}$ and $\kappa^{2}$ see Appendix \ref{ParKL}. Thus, for each block and for each slab of
material (with certain set of phenomenological parameters\cite{Vurga01}), we have an eigenvalue
problem
\be
 \label{eigen}
 \left\{\hat{\bn{\cal H}}_{u,\;l}-E\bn{I}_{2} \right\}\bn{\cal F}_{u,\;l}(z)= \bn{O}_2.
\ee Solving this equation we have
\bea
 \label{block-back}
  \bn{F}(z) = \bn{U}^{\dag}\left[
  \ba{c} \bn{\cal F}_{u}(z) \\
   \bn{\cal F}_{l}(z)
  \ea \right]  =
   \bn{U}^{\dag}
   \sum_{j=1}^{4}\;
  \left( \ba{c}
    \alpha_{j} \;\Bigl( \ba{c} g_{1j} \\ g_{2j} \\ \ea \Bigr) \; \e^{iq_{j}z} \\
    \beta_{j} \;\Bigl( \ba{c} g_{3j} \\ g_{4j} \\ \ea \Bigr) \; \e^{iq_{j}z}
  \ea \right)=\left[
  \ba{c} \bn{F}_{1}(z) \\
   \bn{F}_{2}(z)
  \ea \right] .
\eea The spinor components $g_{ij}$ are given in the Appendix \ref{ParKL}. Here
$\alpha_{j}(\beta_{j})$ stand for the linear combination coefficients of the LI solutions times the
corresponding normalization constant in the ``up''(``low'') subspaces, respectively. We remark that
only two of the values of the momenta $q_{j}$ are LI. They are obtained from the zeros of the
secular fourth-order polynomial determinant of (\ref{eigen}), which is the same for both \textit{u}
and \textit{l} Hamiltonians and then, the energy eigenvalues of the subspace hole states are the
same with different spatial trends. This is perhaps the most important consequence of the
block-diagonalization (B-D) process from (\ref{block}).

When considering the multichannel-multiband transport of \textit{hh} and \textit{lh} through, say,
a III-V semiconductor heterostructures \cite{LPRC02,LPCR04}, one should keep in mind the order of
the basis components to assign correctly the transmission amplitudes. Assuming
$$
 \bn{F}(r)=\bn{F}(z)\cdot\bn{u}_0(r)\mathop{\rm \mbox{{\Large
  e}}}\nolimits^{i\vec{\kappa}\cdot(x\vec{e}_x + y\vec{e}_y)},
$$
we shall choose for the periodic part of the Bloch function
$\bn{u}_0(r)=(u_1,u_2,u_3,u_4)=(\left|\frac{3}{2},\frac{3}{2}\right>,
\left|\frac{3}{2},-\frac{1}{2}\right>, \left|\frac{3}{2},\frac{1}{2}\right>,
\left|\frac{3}{2},-\frac{3}{2}\right>)$, in agreement with the order  $hh_{+3/2}, lh_{-1/2},
lh_{+1/2}, hh_{-3/2}$ given in Ref.\onlinecite{Broido85}. Accordingly, the wave function
$\bn{F}(z)$ must be supplemented by identifying correctly each value of the momenta $q_{j}$ to an
\textit{hh} or \textit{lh} state. To perform this identification we refer to the levels of an
infinite quantum well (iQW) at $\kappa=0$ limit, given by previous calculations
\cite{LCR00,SC1,SP95}. Considering a parabolic approximation to the dispersion law for the
\textit{hh} and \textit{lh} bands in this limit, the energy can be cast as \bea
 \label{escalera}
 E_{hh} & = & \left(\gamma_{1}-2\gamma_{2}\right)q_{3}^{2} \\
 \nonumber
 E_{lh} & = & \left(\gamma_{1}+2\gamma_{2}\right)q_{1}^{2},
\eea
which unambiguously relates the uncoupled levels \textit{hh}(\textit{lh}) to
$q_{3}(q_{1})$, respectively.

Before going forward, notice that both transfer matrices $M_{fd}$ and $T$ relate states in the
quasi-particle representation characterized by a certain set of quantum numbers and describe the
evolution along $z$ of these modes with no mention to their propagation direction. Nevertheless, as
suggested above we can alternatively describe the system using the propagation modes representation
\cite{PPP95}, \textit{i.e.} the representation where right and left moving states are resolved into
separated components. This idea in combination with the mentioned adjustments lead us to express
the wave function (\ref{block-back}) and its derivative in terms of an eight-component state vector
(\ref{SV2}) \be
 \label{FF2}
 \bn{\Psi}(z)= \bn{W}\left\vert \ba{cccc}
                            \bn{\mathbb{G}} & (\bn{\mathbb{G}})^* \\
                        i\bn{\mathbb{G}}\bn{\mathbb Q} & (i\bn{\mathbb{G}})^*\bn{\mathbb{Q}}
                        \ea \right\vert
  \left\vert \ba{ccc}
               \stackrel{\rightarrow}{\bn{\varphi}}(z) & \bn{O}_{4}\\
               \bn{O}_{4} & \stackrel{\leftarrow}{\bn{\varphi}}(z)
   \ea \right\vert \left\vert \ba{c}
                     \bn{a} \\
                     \bn{b}
                   \ea \right\vert,
\ee
\ni where
\beann
 \mbox{\boldmath $W$}=\left[
 \begin{array}{cc}
  \mbox{\boldmath $V$} & \mbox{\boldmath $O$}_4 \\
  \mbox{\boldmath $O$}_4 & \mbox{\boldmath $V$}
 \end{array}
  \right] & , & \mbox{\bm $V$} = \left\vert
 \begin{array}{cccc}
 \mathop{\rm \mbox{{\Large e}}}\nolimits^{i\phi} & 0 & 0 & 0 \\
 0 & \mathop{\rm \mbox{{\Large e}}}\nolimits^{i\eta} & 0 & 0 \\
 0 & 0 & \mathop{\rm \mbox{{\Large e}}}\nolimits^{-i\eta} & 0 \\
 0 & 0 & 0 & \mathop{\rm \mbox{{\Large e}}}\nolimits^{-i\phi}
 \end{array}\right\vert,
\eeann
\ni being
\beann
 \eta = \frac{1}{2} \left\{\arctan(\frac{k_{x}}{k_{y}})- \arctan(\frac{
 2\gamma_{3} k_{x} k_{y}} {\gamma_{2} (k_{y}^{2} - k_{x}^{2})}) \right\}
 & \;\; \mb{and}& \;\;
 \phi= \frac{1}{2} \left\{\arctan(\frac{k_{x}}{k_{y}}) + \arctan(\frac{%
 2\gamma_{3}k_{x}k_{y}} {\gamma_{2}(k_{y}^{2}-k_{x}^{2})}) \right\}.
\eeann \ni The parameters $\eta$ and $\phi$ are taken in the form required by the unitary
transformation\cite{Broido85} $\bn{U}$  for block-diagonalization (\ref{block}). Hereafter
$\gamma_{1}$, $\gamma_{2}$ and $\gamma_{3}$ are the L\"uttinger parameters of the layer.
We have defined
\beann
 \mbox{\boldmath $\mathbb{G}$}= \left\vert
 \begin{array}{rrcc}
  g_{13} & g_{11} & g_{41} & g_{43} \\
  g_{23} & g_{21} & g_{31} & g_{33} \\
  -g_{23}& -g_{21} & g_{31}& g_{33} \\
  -g_{13} & -g_{11} & g_{41} & g_{43}
 \end{array}
 \right\vert & ,\; \mb{and} \; & \mbox{\boldmath $\mathbb{Q}$}= \left\vert
 \begin{array}{cccc}
  q_3 & 0 & 0 & 0 \\
  0 & q_1 & 0 & 0 \\
  0 & 0 & q_1 & 0 \\
  0 & 0 & 0 & q_3
 \end{array}\right\vert.
\eeann In (\ref{FF2}) the $(4 \times 1)$ vectors $\bn{a}(\bn{b})$ contain the corresponding
$a_{j}(b_{j})$ linear combination coefficients of the LI solutions times the corresponding
normalization constant of the configuration space in the ``up''(``low'') subspaces, respectively.
Defining \be \bn{\cal N}=\bn{W}\left\vert \ba{cccc}
                            \bn{\mathbb{G}} & (\bn{\mathbb{G}})^* \\
                        i\bn{\mathbb{G}}\bn{\mathbb Q} & (i\bn{\mathbb{G}})^*\bn{\mathbb{Q}}
                        \ea \right\vert
\ee we have

\be
 \label{FF3}
 \bn{\Psi}(z)= \left\vert \ba{cccc}
                            \bn{\cal N}_{11} & \bn{\cal N}_{12} \\
                        i\bn{\cal N}_{11}\bn{\mathbb{Q}} & (i)^*\bn{\cal N}_{12}\bn{\mathbb{Q}}
                        \ea \right\vert
  \left\vert \ba{ccc}
               \stackrel{\rightarrow}{\bn{\varphi}}(z) & \bn{O}_{4}\\
               \bn{O}_{4} & \stackrel{\leftarrow}{\bn{\varphi}}(z)
   \ea \right\vert \left\vert \ba{c}
                     \bn{a} \\
                     \bn{b}
                   \ea \right\vert = \bn{\cal N}\;\bn{\Phi}(z).
\ee This relation is essential to achieve the purpose posted at the beginning of this section.
Thus, from (\ref{FF3}), the expression (\ref{DefMfd}) that relates $\bn{\Psi}(z)$ at any two points
$z_1$ and $z_2$ of the scatterer system can be written as \be
 \label{FiMfdFi}
 \left\{
 \bn{\cal N}\;\bn{\Phi}(z_{\mb{\tiny 2}}) \right\}_{2}=
 \bn{M}_{fd}(z_{\mb{\tiny 2}},z_{\mb{\tiny 2}})
 \left\{
 \bn{\cal N}\;\bn{\Phi}(z_{\mb{\tiny 1}}) \right\}_{1}.
\ee
Hence the state vectors (\ref{SV2}) in these regions are connected as
\be
 \label{FiNFi}
 \bn{\Phi}(z_{\mb{\tiny 2}})= \bn{\cal N}^{-1}_{2}
 \bn{M}_{fd}(z_{\mb{\tiny 2}},z_{\mb{\tiny 1}})
 \bn{\cal N}_{1}
 \bn{\Phi}(z_{\mb{\tiny 1}}),
\ee
therefore
\begin{equation}
  \mbox{\boldmath $M$}_{sv}(z_{\mb{\tiny 2}},z_{\mb{\tiny 1}})=
  \mbox{\boldmath $\mathcal{N}$}^{-1}_{2}
  \mbox{\boldmath $M$}_{fd}(z_{\mb{\tiny 2}},z_{\mb{\tiny 1}})
  \mbox{\boldmath $\mathcal{N}$}_{1}.
\end{equation}
The convenience of this $(8\times 8)$ matrix identity is likely found in the clear
advantages of using both types of transfer matrices for the description of tunneling and
related properties of particles moving through multilayered structures. This result
becomes a profitable platform \cite{LPRC02,LPCR04} to circumvent the usual assumptions and
restrictions that the current descriptions \cite{Xia88,Chu89,WA89,ChaoChu91,SP95,KCR97}
face in describing the hole tunnelling in heterostructures.

We emphasize several points here: Attention must be paid to orto-normalization of the spinor
components $g_{ij}$, content in the $(8 \times 8)$ transformation matrix $\bn{\cal N}$, to prevent
prompt failures during numerical simulation of quantum transport phenomena of heavy and light holes
through heterostructures. For systems well-described within the EFA the orto-normalization
procedure is not a trivial question and differs from routine process, basically, by the presence of
linear $\hat{k}_{z}$ terms in (\ref{Hu},\ref{Hl}). A detailed study on that subject is published
elsewhere\cite{Perez04,Teje01}.

\subsection{TRI and SII symmetries for the $\protect\kappa \neq 0$ case in the
            $(2 \times 2)$ KL model}

Let us analyze the implications of the TRI and SII symmetries (in the $(4 \times 4)$ KL model) on
some physical quantities defined in the $(2 \times 2)$ subspaces. We shall start with the TRI
symmetry. To exhibit this implications on the time reversal operator operator we will use its
representation  reported by Pasquarello {\it et al.} in Ref.\onlinecite{APB}. Before
block-diagonalizing the time reversal and space inversion operators, we need to transform them from
the Pasquarello's basis representation to the Broido and Sham basis representation \cite{Broido85}.
This transformation is done by means of
\begin{eqnarray}
 \label{Andre2Broido}
 \mbox{\boldmath $R$} & = & \left\vert
\begin{array}{cccc}
-1 & 0 & 0 & 0 \\
0 & 0 & 1 & 0 \\
0 & i & 0 & 0 \\
0 & 0 & 0 & -i
\end{array}
\right\vert\;,
\end{eqnarray}
\noindent In the following, $\sigma_x, \sigma_y, \sigma_z$ are the three Pauli matrices and the
coordinate origin will be placed in the inversion symmetry point ($z = 0$).

\subsubsection{Time Reversal Invariance}

\noindent The $(4\times 4)$ time reversal operator in the Broido and Sham basis is given by
\begin{equation}
\label{APB2BS-OpTR} \hat{T}_{\mb{\tiny BS}}=\mbox{\boldmath $K$}_{\mb{\tiny
BS}}\hat{C}=\mbox{\boldmath $R$}\mbox{\boldmath $K$}_{\mb{\tiny PB}}\hat{C}\mbox{\boldmath
$R$}^{-1}= \mbox{\boldmath $R$} \left[
\begin{array}{cc}
\mbox{\boldmath $O$}_2 & \mbox{\boldmath $\sigma$}_{y} \\
\mbox{\boldmath $\sigma$}_{y} & \mbox{\boldmath $O$}_2
\end{array}
\right]\mbox{\boldmath $R$}^{T}\hat{C}= \left[
\begin{array}{cc}
\mbox{\boldmath $O$}_2 & \mbox{\boldmath $\sigma$}_{x} \\
-\mbox{\boldmath $\sigma$}_{x} & \mbox{\boldmath $O$}_2
\end{array}
\right]\hat{C}.
\end{equation}
Hereafter subindexes BS and PB stand for Broido-Sham and Pasquarello-Bassani,
respectively. After applying the transformation \cite{Broido85}
\be
\bn{U}
=\frac{1}{\sqrt{2}} \left\vert
 \begin{array}{cccc}
 \mathop{\rm \mbox{{\Large e}}}\nolimits^{-i\phi} & 0 & 0 & -\mathop{\rm \mbox{{\Large e}}}\nolimits^{i\phi}\\
 0 & \mathop{\rm \mbox{{\Large e}}}\nolimits^{-i\eta} & -\mathop{\rm \mbox{{\Large e}}}\nolimits^{i\eta} & 0 \\
 0 & \mathop{\rm \mbox{{\Large e}}}\nolimits^{-i\eta} & \mathop{\rm \mbox{{\Large e}}}\nolimits^{i\eta} & 0 \\
 \mathop{\rm \mbox{{\Large e}}}\nolimits^{-i\phi}  & 0 & 0 & \mathop{\rm \mbox{{\Large e}}}\nolimits^{i\phi}
 \end{array}\right\vert,
 \ee
the time reversal operator $\hat{T}_{\mb{\tiny BS}}$ becomes

\be {\hat{\cal{T}}_{\mb{\tiny BS}}}=\bn{URK}_{\mb{\tiny PB}}\bn{R}^T\bn{U}^T\hat{C}=
\left[\begin{array}{cc}
\mbox{\boldmath $O$}_2 & \mbox{\boldmath $\sigma$}_{x} \\
-\mbox{\boldmath $\sigma$}_{x} & \mbox{\boldmath $O$}_2
\end{array}
\right]\hat{C}. \ee As a consequence the TRI requirement implies for the $(2 \times 2)$ subspace
Hamiltonians the fulfillment of the following condition
\begin{equation}
 \label{Hu2Hl-TRI}
\left.
\begin{array}{ccc}
\mbox{\bm $\sigma$}_{x}\hat{
\mbox{\boldmath $\cal H$}}^*_{u}(z)\mbox{\bm $\sigma$}_{x} & = & \hat{\mbox{\boldmath $\cal H$}}_{l}(z) \\
\mbox{\bm $\sigma$}_{x}\hat{ \mbox{\boldmath $\cal H$}}^*_{l}(z)\mbox{\bm $\sigma$}_{x} & = &
\hat{\mbox{\boldmath $\cal H$}}_{u}(z)
\end{array}
\right\}.
\end{equation}
It is then clear that the TRI of the KL Hamiltonian in the $(4\times4)$ space does not
imply the TRI of the Hamiltonians in the $(2\times2)$ subspaces. It is easy to show also
that
\begin{equation}
 \label{TRI4Fu}
 \mbox{\bm $\sigma$} _{x}\mbox{\bm $F$}^*_{u,\;l}(z) =
 \mbox{\bm $F$}_{l,\;u}(z).
\end{equation}
and hence
\begin{equation}
 \label{Mu2Ml-TRI}
 \begin{array}{ccc}
 \mbox{\bm $\Gamma$}_{x}
 \mbox{\boldmath $M$}^*_{u,\;l}(z)\mbox{\bm $\Gamma$}_{x} & = & \mbox{\boldmath $M$}_{l,\;u}(z)
 \end{array},
\end{equation}
here:
$$
 \bn{\Gamma}_{x}=\left\vert
 \begin{array}{cc}
  \bn{\sigma} _{x} & \bn{O}_2 \\
  \bn{O}_2 & \bn{\sigma}_{x}
 \end{array}\right\vert.
$$

As can be seen from previous relations, the considered physical quantities of the
subspaces \textit{u} and \textit{l} should be related to each other in order to satisfy
the TRI requirement in the $(4 \times 4)$ space.

\subsubsection{Space Inversion}

It is convenient to emphasize that the space inversion symmetry we are dealing with is the
one which changes only the sign to the coordinate perpendicular to the interfaces of the
Q2D system. The usual space inversion \cite{APB} makes $\{x,y,z\}\rightarrow\{-x,-y,-z\}$
then, when taking 2D-Fourier transform in the $[x,y]$ plane, change the signs of $k_{x}$
and $k_{y}$ also.

As in the case of time reversal, we need  to know first the space inversion operator (\ref
{OpSI}) for the $(4 \times 4)$ KL space in the Broido and Sham basis\cite{Broido85}. This
is obtained from

\begin{equation}
 \label{APB2BS-OpSI}
 {\hat{\cal{S}}}_{\mb{\tiny BS}}=\mbox{\boldmath $R$}{\hat{\cal{S}}}_{\mb{\tiny PB}}
 \mbox{\boldmath
$R$}^{-1}=\mbox{\boldmath $R$}\left[
 \begin{array}{cc}
 -\mbox{\boldmath $\sigma$}_{z} & \mbox{\boldmath $O$}_2 \\
\mbox{\boldmath $O$}_2 & -\mbox{\boldmath $\sigma$}_{z}
\end{array}
\right]\mbox{\boldmath $R$}^{-1} = \left\vert
\begin{tabular}{rr}
$-\mbox{\boldmath $I$}_{2}$ & $\mbox{\boldmath $O$}_{2}$ \\
$\mbox{\boldmath $O$}_{2}$ & $\mbox{\boldmath $I$}_{2}$%
\end{tabular}
\right\vert.
\end{equation}

It is easy to verify that the $(4 \times 1)$ wave function $\bn{F}(z)$ defined in
(\ref{block-back}), under this operator action, satisfies the symmetry condition
(\ref{OpSI}). It can be seen also that the time reversal and space inversion operators
satisfy $\{\hat{\cal{T}}_{\mb{\tiny BS}}, {\hat{\cal{S}}}_{\mb{\tiny BS}}\} = \hat{0}$,
\textit{i.e.} they anti-commute. This leads to the existence of an additional symmetry
operator ${\hat{\cal{S}}}_{\mb{\tiny BS}}\hat{\cal{T}}_{\mb{\tiny BS}}$ that changes a
wave function for a given $\kappa$ with a definite parity into the same function of
$\kappa$ but with opposite parity \cite{APB,LCR00}. The invariance of the wave function
$\bn{F}(z)$ in (\ref{block-back}) under space inversion operation (\ref {OpSI}) implies
\[
{\hat{\cal{S}}}_{\mb{\tiny BS}}\left[
\begin{tabular}{l}
$\bn{F}_{1}(z)$ \\
$\bn{F}_{2}(z)$
\end{tabular}
\right] = p\left[
\begin{tabular}{c}
$-\bn{F}_{1}(-z)$ \\
$\;\;\;\bn{F}_{2}(-z)$
\end{tabular}
\right]=\left[
\begin{tabular}{l}
$\bn{F}_{1}(z)$ \\
$\bn{F}_{2}(z)$
\end{tabular}
\right].
\]

Since
\begin{equation}
 \label{OpSI2}
\mbox{\boldmath $U$}{\hat{\cal{S}}}_{\mb{\tiny BS}}\mbox{\boldmath $U$}^\dag = \left[
\begin{array}{cc}
\mbox{\boldmath $O$}_2 & \mbox{\boldmath $\sigma$}_{x} \\
\mbox{\boldmath $\sigma$}_{x} & \mbox{\boldmath $O$}_2
\end{array}
\right],
\end{equation}
the space inversion invariance of the KL Hamiltonian leads us to
\be
 \left\vert
  \begin{array}{cc}
   \bn{O}_{2} & \bn{\sigma}_{x} \\
   \bn{\sigma}_{x} & \bn{O}_{2}
  \end{array} \right\vert
   \left\vert \ba{cc}
  \hat{\bn{\cal H}}_{u}(-z) & \bn{O}_{2} \\
  \bn{O}_{2} & \hat{\bn{\cal H}}_{l}(-z)
 \ea \right\vert
  \left\vert
  \begin{array}{cc}
   \bn{O}_{2} & \bn{\sigma}_{x} \\
   \bn{\sigma}_{x} & \bn{O}_{2}
  \end{array} \right\vert =
  \left\vert \ba{cc}
  \hat{\bn{\cal H}}_{u}(z) & \bn{O}_{2} \\
  \bn{O}_{2} & \hat{\bn{\cal H}}_{l}(z)
 \ea \right\vert.
\ee Therefore
\begin{equation}
 \label{Hu2Hl-SI}
\left.
\begin{array}{ccc}
\hat{\mbox{\boldmath $\cal H$}}_{u}(z) & = & \mbox{\bm $\sigma$}_{x}
\hat{\mbox{\boldmath $\cal H$}}_{l}(-z) \mbox{\bm $\sigma$}_{x} \\
\hat{\mbox{\boldmath $\cal H$}}_{l}(z) & = & \mbox{\bm $\sigma$}_{x}
\hat{\mbox{\boldmath$\cal H$}}_{u}(-z) \mbox{\bm $\sigma$}_{x}
\end{array}
\right\}.
\end{equation}
\noindent For the transfer matrices of the first kind and for the envelope wave functions we
have
\begin{equation}
 \label{Mu2Ml-SI}
\begin{array}{ccc}
\mbox{\boldmath $M$}_{u,\;l}(z) & = & \mbox{\bm $\Gamma$}_{x}\mathbf{M}%
_{l,\;u}(-z)\mbox{\bm $\Gamma$}_{x}
\end{array},
\end{equation}
\begin{equation}
 \label{SI4Fu}
 \mbox{\bm $\sigma$}_{x}\mbox{\bm $F$}_{u,\;l}(-z)=p\mbox{\bm $F$}_{l,\;u}(z).
\end{equation}

Here again the transfer matrices and wave functions, appearing in the first and second
members of equations (\ref{Mu2Ml-SI}) and (\ref{SI4Fu}), belong to different subspaces.
For $\kappa \neq 0$ the space inversion symmetry in the $(4\times 4)$ space does not imply
space inversion symmetry in the $(2\times 2)$ spaces. Hence, the wave functions in both
subspaces does not have a definite parity.

The mixing of $hh$ with $lh$ states increases as we go away from the center of the
Brillouin Zone, i.e., as $\kappa$ differs from $0$. The square modulus of the wave
function $(\mathcal{D}_{u,\;l})$ was evaluated by using the representation for the TM of
\textit{first kind} for the hole state. Taking advantage of (\ref{SI4Fu}) the following
useful relation is obtained:
\begin{equation}
\label{Fu-Fl}
\mathcal{D}_{l}(E,\kappa,z) = \mathcal{D}_{u}(E,\kappa,-z).
\end{equation}
where $\mathcal{D}_{u,\;l}$ are defined as
\begin{equation}
 \label{|Fu|^2}
\mathcal{D}_{u,\;l}(E,\kappa,z) = \left[\mathbf{M}^{u,\;l}_{12}(z,z_{o})\bn{\cal
F}_{u,\;l}(z_{o})\right]^{\dagger} \left[\bn{M}^{u,\;l}_{12}(z,z_{o})\bn{\cal
F}_{u,\;l}^{\,\prime}(z_{o})\right],
\end{equation}
\ni where $\mathbf{M}^{u,\;l}_{12}(z,z_{o})$ are blocks of the TM of \textit{first kind}
$\mathbf{M}^{u,\;l}$, respectively.

\subsection{TRI and SII symmetries for the $\protect\kappa = 0$ case in the
            $(2 \times 2)$ KL model}

At $\kappa = 0$ the hole states decouple in the case of no external field or no strain,
given independent series of levels (\ref{escalera}) for the infinite quantum well boundary
conditions. Both $\hat{\bn{\cal H}}_{u}$ and $\hat{\bn{\cal H}}_{l}$ are diagonal and thus
invariant under time reversal and space inversion operations. The time reversal operator
in this case is $\bna{\sigma}_z\hat{C}$, which commute with the Hamiltonian, and the wave
functions satisfy
\begin{equation}
 \label{Fu-Fu}
 \mbox{\boldmath $\sigma$}_{z}\mbox{\boldmath $F$}_{u,\;l}(-z) = p\;
 \mbox{\boldmath $F$}_{u,\;l}(z),
\end{equation}
\noindent then the \textit{hh} and \textit{lh} components of the vectors $\bn{F}_{u,\;l}(z)$ posses
definite but opposite parities. At the $\Gamma$ point the \textit{up} and \textit{low} subspaces
satisfy FC, TRI and SII and belong to the \emph{symplectic universality class} \cite {PPP95}, as
the original $(4 \times 4)$ space of solutions.

\section{Conclusions}

We establish the relation between the TM of the first kind (frequently used to determine
the energy spectra) and the TM of the second kind (widely used in the scattering approach)
in general and for the specific case of KL Hamiltonian model. This relation allows us to
take the advantage of using the appropriate transfer matrix according to the circumstances
one is dealing with. It is possible to avoid the usual assumptions and restrictions that
the current descriptions face in describing tunnelling of holes
\cite{Xia88,Chu89,WA89,ChaoChu91,SP95,KCR97}.

We deduce new symmetry requirements on the TM of the \textit{first kind} $\bn{M}_{fd}$ and
SII requirements on TM of the \textit{second kind} $\bn{M}_{sv}$ for $N>1$ case of actual
interest in quantum-transport problem. Any EFA model having a master equation like that of
the References \onlinecite{Malik99,RPA01c} are expected to have the same or similar
properties. In fact, a great part of the analysis carried out in sections \ref{Def} and
\ref {Sym} can be extended to those problems with minor changes.

We made clear that the symmetry analysis in the reduced $(2 \times 2)$ subspaces of the KL
model demand attention to prevent erroneous statements\cite{SP95}. We show that for
$\kappa \neq 0$ the time reversal and space inversion invariance requirements of the $(4
\times 4)$ KL Hamiltonian does not imply the invariance of the `up" and ``low"
Hamiltonians and related objects, separately. However, to fulfill the whole Hamiltonian
invariance they have to transform one to the other according to the corresponding
expressions obtained here. It has been also deduced that for $\kappa =0$, the ``up" and
``low" subspaces recover continuously the time reversal and space inversion symmetries, as
expected.

Besides the basic interest of this topic, the relations obtained here are useful and convenient in
controlling bothersome numerical calculations, as well as in studying the relevant quantities in
the multichannel-multiband tunnelling of holes within a more realistic approach. Expressions of
section \ref{Sym} were verified numerically within computer uncertainty.

\section*{Acknowledgments}

%\ack
We are indebted to F. Garcia-Moliner and F. de Le\'{o}n-P\'{e}rez and Herbert Simanjuntak the
fruitful discussions with some of the authors. This work has been performed under a CONACyT(Mexico)
grand No.E120.1781 and received support from Departamento de Ciencias of UAM-A, Mexico and from
C\'{a}tedra de Ciencia Contempor\'{a}nea of ``Jaume I'' University, Spain.

%\section*{Appendix}
\appendix

\section{Parameters of the $(2 \times 2)\;\; \hat{\mbox{\boldmath $\cal H$}}
_{u}$ and $\hat{\mbox{\boldmath $\cal H$}}_{l}$ KL Hamiltonians}

\label{ParKL} \setcounter{section}{1}
\begin{equation}
\label{ParKL1} \left.
\begin{array}{cc}
k_{\pm} = k_{x} \pm i k_{y}\mbox{;} & \kappa^{2} = k_{x}^{2} + k_{y}^{2} \\
A_{1} = \frac{\hbar^{2}}{2m_{0}} (\gamma_{1}+\gamma_{2})\mbox{;} & A_{2} =
\frac{\hbar^{2}}{2m_{0}} (\gamma_{1}-\gamma_{2}) \\
B_{1} = \frac{\hbar^{2}}{2m_{0}} (\gamma_{1}+2\gamma_{2})\mbox{;} & B_{2} =
\frac{\hbar^{2}}{2m_{0}} (\gamma_{1}-2\gamma_{2}) \\
C_{xy} = \sqrt{3}\;\frac{\hbar^{2}}{2m_{0}}\;\sqrt{\gamma_{2}^{2}
(k_{x}^{2}-k_{y}^{2})^{2}+4\gamma_{3}^{2}k_{x}^{2}k_{y}^{2}}\mbox{;} &
D_{xy} = \sqrt{3}\;\frac{\hbar^{2}}{m_{0}}\;\gamma_{3}\;\kappa
\end{array}
\right\}
\end{equation}
The non-ortonormalized spinor components $g_{ij}$
\begin{equation}
\left.{\hspace{1mm}}
\begin{tabular}{ccc}
$g_{11}$ & = & $A_{2}\kappa^{2}+B_{1}q_{1}^{2}+V(z) $ \\
$g_{13}$ & = & $A_{2}\kappa^{2}+B_{1}q_{3}^{2}+V(z) $ \\
$g_{22}$ & = & $-(C_{xy}+iD_{xy}q_{2}) $ \\
$g_{24}$ & = & $-(C_{xy}+iD_{xy}q_{4})$ \\
$g_{31}$ & = & $A_{1}\kappa^{2}+B_{2}q_{1}^{2}+V(z) $ \\
$g_{33}$ & = & $A_{1}\kappa^{2}+B_{2}q_{3}^{2}+V(z)$ \\
$g_{42}$ & = & $-(C_{xy}+iD_{xy}q_{2}) $ \\
$g_{44}$ & = & $-(C_{xy}+iD_{xy}q_{4}) $
\end{tabular}
\right\},
\end{equation}

\noindent from $\hat{\mbox{\boldmath $\cal H$}}_{u}$ and $\hat{%
\mbox{\boldmath $\cal H$}}_{l}$ peculiarities follow
\begin{equation}
\left.{\hspace{1mm}}
\begin{tabular}{ccc}
$q_{2}$ & = & $-q_{1}$ \\
$q_{4}$ & = & $-q_{3}$ \\
$g_{11}$ & = & $g_{12}$ \\
$g_{13}$ & = & $g_{14}$ \\
$g_{22}$ & = & $g_{21}^{*}$ \\
$g_{24}$ & = & $g_{23}^{*}$ \\
$g_{31}$ & = & $g_{32}$ \\
$g_{33}$ & = & $g_{34}$ \\
$g_{42}$ & = & $g_{41}^{*}$ \\
$g_{44}$ & = & $g_{43}^{*}$%
\end{tabular}
\right\},
\end{equation}
\noindent being
\begin{eqnarray*}
\left.{\hspace{1mm}}
\begin{tabular}{ccc}
$g_{1j},g_{3j}$ & real &  \\
$g_{2j},g_{4j}$ & complex &
\end{tabular}
\right\}.
\end{eqnarray*}
\noindent The L\"{u}ttinger parameters $\gamma_{1},\gamma_{2},\gamma_{3}$
typify each layer of the heterostructure.

\section{Transformation symmetry matrices}

\label{MatTran} Given the transformation between the matrices $%
\mbox{\boldmath $M$}_{fd}(z,z_{0})$ and $\mbox{\boldmath $M$}_{sv}(z,z_{0})$%
, it is easy to see that the FC constrain on $\mbox{\boldmath $M$}%
_{sv}(z,z_{0})$ (Sec.\ref{Sym}) implies that
\begin{equation}
\mbox{\boldmath $\Sigma$}_{z}=\left( \mbox{\boldmath $\mathcal{N}$}^{\;\dag
}\right)^{-1}\mbox{\boldmath $J$}_{fd} \mbox{\boldmath $\mathcal{N}$}.
\end{equation}
\noindent However, it is worth noticing that, whenever the basis of linearly independent
KL Hamiltonian eigenvectors, is not an orthogonal basis, we will have
\[
\mbox{\boldmath $J$}_{fd}= \left[
\begin{array}{ccc}
 \mbox{\boldmath $u$}^{\dag }(z)-\mbox{\boldmath $u$}(z) & \;\; &
 -\mbox{\boldmath$v$}(z) \\
 \mbox{\boldmath $v$}(z) & \;\; & \mbox{\boldmath $O$}_4
\end{array}
\right].
\]
For TRI and SII requirements on the TM $\mbox{\boldmath $M$}_{sv}(z,z_{0})$
(Sec.\ref{Sym}) we used the matrices $\mbox{\boldmath $\Sigma$}_{x}$ and $%
\mbox{\boldmath $S$}_{sv}$, respectively, which read
\begin{equation}
\mbox{\boldmath $\Sigma$}_{x}= \mbox{\boldmath $\mathcal{N}$}^{-1}%
\mbox{\boldmath $\Sigma$}^{-1}\mbox{\boldmath $\mathcal{N}$}^{*},\;\;\; %
\mbox{\boldmath $S$}_{sv}= \mbox{\boldmath $\mathcal{N}$}\mbox{\boldmath $S$}%
\mbox{\boldmath $\mathcal{N}$}^{-1}.
\end{equation}

\end{document}